\documentclass[11pt]{article}

\usepackage{amssymb}
\usepackage{graphics,graphicx}
\usepackage{citesort}

\begin{document}
\thispagestyle{empty}

\begin{flushright}
                LPT Orsay \\
                September 2003  \\
		hep-th/0309267
                
\end{flushright}
\bigskip

\begin{center}

{\LARGE\bf\sf Classical Liouville action }

\vskip 3mm

{\LARGE\bf\sf  on the sphere with three hyperbolic singularities }

\end{center}
\bigskip

\begin{center}
    {\large\bf\sf
    Leszek Hadasz}\footnote{e-mail: hadasz@th.if.uj.edu.pl} \\
\vskip 1mm
    Laboratoire de Physique Th\`{e}orique, B{\^a}t. 210,\\
    Universit{\'e} Paris-Sud,  91405 Orsay, France \\
        {\em and }\\
    M. Smoluchowski Institute of Physics, \\
    Jagiellonian University,
    Reymonta 4, 30-059 Krak\'ow, Poland \\
\vskip 5mm
    {\large\bf\sf
    Zbigniew Jask\'{o}lski}\footnote{e-mail: Z.Jaskolski@if.uz.zgora.pl
}\\
\vskip 1mm
    Physics Institute\\
    University of Zielona G\'{o}ra\\
    ul. Szafrana 4a,
    65-069 Zielona G\'{o}ra, Poland
\end{center}

\vskip .5cm

\begin{abstract}
The classical solution to the Liouville equation 
in the case of three hyperbolic singularities of its energy-momentum tensor 
is derived and analyzed.
The recently proposed classical Liouville action is 
explicitly calculated 
in this case. 
The result agrees with the classical limit 
of the three point function in the DOZZ solution of the 
quantum Liouville theory. 
\end{abstract}

\newpage
\setcounter{page}{1}
\vskip 1cm
\section{Introduction}
In the last few years a considerable progress in the Liouville theory has been achieved \cite{Teschner:2001rv}.
The solution to the quantum theory based on the structure constants 
proposed by Otto and Dorn~\cite{Dorn:1994xn} 
and by A. and Al. Zamolodchikov \cite{Zamolodchikov:1995aa}
was recently completed by  Ponsot and Teschner~\cite{Ponsot:1999uf,Ponsot:2000mt,Ponsot:2001ng}
and by Teschner \cite{Teschner:2001rv,Teschner:2003en}. 

On the other hand there exists so called geometric approach 
originally proposed by Polyakov \cite{Polyakov82} and further 
developed by Takhtajan 
\cite{Takhtajan:yk,Takhtajan:1994vt,Takhtajan:1993vt,Takhtajan:zi}.
In contrast to the operator formulation of the DOZZ theory
the correlators of primary fields with elliptic and 
parabolic weights are expressed in terms of a path integral 
over conformal class of Riemannian metrics
with prescribed  singularities at the punctures.
Some of the predictions derived from this representation 
can be rigorously proved and lead to deep geometrical results
\cite{ZoTa1,ZoTa2,Zograf,Cantini:2001wr,Takhtajan:2001uj}.

In spite of these achievements 
 the relation between both approaches is not fully
understood~\cite{Takhtajan:1995fd,Teschner:2003at}. This in particular concerns 
the problem
of path integral representation of the Liouville correlators 
with hyperbolic weights which in order requires an appropriate 
choice of the Liouville action functional.
A step in this direction was done 
in our previous paper \cite{Hadasz:2003kp} where we constructed the classical Liouville action
satisfying the Polyakov conjecture in the case
of hyperbolic singularities. 

In the present paper we calculate the classical Liouville action
for three hyperbolic singularities and show that the result agrees with the classical
limit of the corresponding DOZZ three point function. In the case of
three singularities the 
monodromy problem considerably simplifies and the solution can be find in 
terms of hypergeometric functions \cite{Bilal:Gervais}. It
shares all  the properties required 
by the construction of the Liouville action and the proof of the Polyakov
conjecture given in \cite{Hadasz:2003kp}.

The organization of the paper is as follows.
In Sect.2 we briefly describe the $SL(2,\mathbb{R})$ monodromy problem 
for $n\geq 3$ hyperbolic singularities on the Riemann sphere. 
The singularity
structure of the corresponding solutions to the Liouville equation is analyzed.
In Sect.3 we present the essential steps of 
 the construction of the 
classical Liouville action and calculate its partial derivatives with respect to
 the conformal weights.
In Sect.4 we derive  the solution to the monodromy problem in the 
case of three hyperbolic singularities. The corresponding 
solution to the Liouville equation
defines a singular hyperbolic geometry on the sphere. We illustrate the
structure of  the closed geodesics and the singular lines of this geometry
on two examples.
In Sect.5 we find an explicit expression 
for the classical Liouville action and compare it 
with the classical 
limit of the DOZZ three point function 
invariant 
with respect to the reflections $\alpha \to Q-\alpha $.

The classical Liouville action proposed in \cite{Hadasz:2003kp}
can be easily generalized 
to the case of an arbitrary number of elliptic, parabolic and hyperbolic
singularities on the Riemann sphere. 
The results of the present paper provide additional justification for this 
construction. The path integral representation
of correlation functions based on this action functional  may provide a new
insight into the geometric content of the quantum Liouville theory
and is certainly worth further investigations.
Also the classical theory itself  rises some 
 interesting questions concerning
 the existence and the uniqueness of regular solutions
to the $SL(2,\mathbb{R})$ monodromy problem. Let us finally mention that
the classical solutions with hyperbolic singularities are 
of  interest in 3-dim gravity where they describe multi-black-hole 
configurations \cite{Banados:wn,Brill:1995jv,Welling:1997fw,Cantini:2001im,Cantini:2002jw}. 
\section{$SL(2,\mathbb{R})$ monodromy problem }

The construction of the Liouville action proposed in
\cite{Hadasz:2003kp} relies on solutions of 
   a special monodromy 
problem for the Fuchsian equation 
\begin{equation}
\label{Fuchs}
\partial^2\psi(z) + \frac{1}{8\beta}T(z)\psi(z) = 0 
\end{equation}
where
\begin{equation}
\label{T:zz}
T(z) = \sum\limits_{j=1}^{n}\left[\frac{\Delta_j}{(z-z_j)^2} +
\frac{c_j}{z-z_j}\right]
\end{equation}
is assumed to be regular at infinity and $\Delta_j$ are
hyperbolic conformal weights
\begin{equation}
\label{Delty:geom}
\Delta_j = 2\beta(1+\lambda_j^2)\ ,
\hskip 5mm
\lambda_j\in \mathbb{R}\ ,
\hskip 3mm 
j=1,...,n\ .
\end{equation}
The problem is to adjust the accessory parameters
$c_j$ in such a way  that the equation (\ref{Fuchs}) admits a
fundamental system  
of solutions with $SL(2,\mathbb{R})$ monodromies around all singularities. 
If
$
\Psi(z) 
$
is such a system with Wronskian normalized to 1, 
then for each singularity $z_j$ there exists
an element $B_j\in SL(2,\mathbb{R})$ such that  the system $\Psi^j(z) = 
B_j\,
\Psi(z)$ 
has the canonical form
\begin{equation}
\label{psis}
\Psi^j(z) =\left(
\!\!
\begin{array}{c}
\psi^j_+(z)\\
\psi^j_-(z)
\end{array}
\!\!\right)
\;=\;
(i\lambda_j)^{-{1\over 2}}
\left(
\!\!
\begin{array}{c}
{\rm e}^{ {i\over 2}\vartheta_j} 
(z-z_j)^{\frac{1}{2}(1+ i\lambda_j)} u^j_+ (z)\\
{\rm e}^{ -{i\over 2}\vartheta_j} 
(z-z_j)^{\frac{1}{2}(1- i\lambda_j)} u^j_- (z)
\end{array}
\!\!\right)
 \ ,
\end{equation}
where $ \vartheta_j \in \mathbb{R}.$
$ u^j_\pm(z)$ are  analytic functions
on the disc $D_j=\{z\in X: |z-z_j| < \min\limits_{i,\ i\neq j} |z_i-z_j|\}$
such that $u^j_\pm(0) =1.$ 
Each system $\Psi^j(z)$ defines by the formula 
\begin{equation}
\label{cliouville}
e^{-{\varphi(z,\bar z)\over 2}} \; = \;\pm\;\frac{i\sqrt{m}}{2}
\left(
\psi^j_+(z)\overline{\psi^j_-(z)}
-\overline{\psi^j_+(z)}\psi^j_-(z)
\right)
\end{equation}
the same single-valued, real solution $\varphi(z,\bar z)$ to the
Liouville equation 
\begin{equation}
\label{liouville}
\partial\bar\partial \varphi
={\textstyle \frac{m}{2}}{\rm e}^\varphi \ .
\end{equation}
In the case of hyperbolic weights  the singularity structure of $\varphi$
 can be described by means of the multi-valued  conformal maps
\begin{equation}
\label{rho}
\rho_j(z)
=
\left(\frac{\psi^j_+(z)}{\psi^j_-(z)}\right)^{\frac{1}{i\lambda_j}} 
 \; = \; 
{\rm e}^{\vartheta_j\over \lambda_j}
\left[z-z_j + \frac{c_j}{2\Delta_j}(z-z_j)^2 +
{\cal O}\left((z-z_j)^3\right)\right] \ .
\end{equation}
Singularities of 
$
\rho_j(z)$ on $X =
\hat{\mathbb C}\setminus \bigcup_{j=1}^n\{z_j\}
$ 
are branch points of infinite
order located at zeros of $\psi^j_+(z)$ or $\psi^j_-(z)$.
In spite of this complicated analytical structure 
one easily checks that  the metric
${\rm e}^{\varphi}d^2z$ coincides with the pull-back of the metric
\begin{equation}
\label{Seiberg}
\frac{\lambda^2_jd^2\rho}{\,m\,|\rho|^2 \sin^2\left(\lambda_j\log
|\rho|\right)}
\end{equation} 
by the map $\rho_j(z)$. 
We shall briefly discus the consequences of this fact.

\newpage

The only singularities of the metric  ${\rm e}^{\varphi}d^2z$
are closed, non-intersecting smooth lines with the following properties:
\begin{itemize}
\item the locations $z_j$ of the energy-momentum tensor singularities
 are  the only limit points\footnote{with respect to the flat metric on the complex plane} of these lines;
\item
there are no regions in $\hat\mathbb{C}$
bounded by singular lines which do not contain at least one point $z_j$;
\item
none of the  singular lines 
which separate $z_j$ from all other  points $z_i, i\neq j$ 
contains  branch points of $\rho_j$.
\end{itemize}
Let $\Sigma_j$ be the ``most distant from $z_j$'' singular line around $z_j,$ 
defined by the property that there are no singular lines separating
$\Sigma_j$ from 
all the  points $z_i, i\neq j$. $\Sigma_j$ is an inverse image of one of
the singular 
lines of the metric (\ref{Seiberg}) on the $\rho$ plane: 
$$
\rho_j\left(\Sigma_j \right)=  \{ \rho\in \mathbb{C} : \lambda_j \log
|\rho|=\pi l_j\} 
$$
for some $l_j\in \mathbb{Z}$. It follows from the behavior of the hyperbolic geometry 
near singular lines that there exist a closed geodesic $\Gamma_j$ separating
$\Sigma_j$ from all other ``most distant'' singular lines $\Sigma_i, i\neq j$. 
Then
$$
\Gamma_j = \rho_j^{-1} \left( \{ \rho \in \mathbb{C}: \lambda_j \log|\rho|=\pi(l_j+{\textstyle {1\over 2} })\} \right) 
$$
and the map $\rho_j$ is invertible on the {\it hole $H_j$ around $z_j$}:
$$
H_j= \rho_j^{-1} \left( \{ \rho \in \mathbb{C}: \lambda_j \log|\rho|\leq \pi(l_j+{\textstyle {1\over 2} })\} \right) 
$$
The geometry on $H_j$ is therefore isomorphic to that of
the metric (\ref{Seiberg}) on the disc 
$\{ \rho \in \mathbb{C}: \lambda_j \log|\rho|\leq \pi(l_j+{\textstyle {1\over 2} })\}$.

The region ``between the holes'' $(H_i\cap H_j=\emptyset$ for all $i\neq j)$:
\begin{equation}
\label{M}
M  \equiv \hat{\mathbb C} \setminus \bigcup\limits_{j=1}^n H_j 
\end{equation}
carries a  hyperbolic geometry with geodesic boundaries.
It is however not guaranteed that for $n>3$ 
there are no line singularities in $M$.
We say that a solution to the $SL(2,\mathbb{R})$ monodromy problem is {\it regular}
if the corresponding metric ${\rm e}^{\varphi}d^2z$ is regular on $M$.
Note that in the case of three singularities any solution is regular.

\section{Classical Liouville action, $n\geq 3$}

Each regular solution to the $SL(2,\mathbb{R})$ 
monodromy problem uniquely determines the region $M\subset \hat \mathbb{C}$ (\ref{M}).
We start from
the standard Liouville action on $M$
\begin{equation}
\label{the:action}
S_{\rm\scriptscriptstyle L}\left[M,\phi\right]
\;  = \;
\frac{1}{2\pi}\int\limits_M\! d^2z\;
    \left(\partial\phi\bar\partial\phi + m\,{\rm e}^\phi\right)
    + \frac{1}{2\pi}\int\limits_{\partial M}\!|dz|\;\kappa_z\phi\ ,
\end{equation}
where $d^2z = \frac{i}{2}dz\wedge \bar dz$
and $\kappa_z$ is a geodesic curvature of $\partial M$  computed in the
flat metric on the complex plane.
For finite locations $z_j$ of hyperbolic singularities $M$ is unbounded 
and one has to impose an appropriate asymptotic conditions
on admissible solutions. It can be done by means of the modified action
\begin{eqnarray}
\label{the:actionR}
S^\infty_{\rm\scriptscriptstyle L}\left[M,\phi\right] &=&
\lim_{R\to\infty} S^R_{\rm\scriptscriptstyle L}\left[M,\phi\right] \ ,\\
\nonumber
S^R_{\rm\scriptscriptstyle L}\left[M,\phi\right]
&=&
\frac{1}{2\pi}\int\limits_{M^R}\! d^2z\;
    \left(\partial\phi\bar\partial\phi + m\,{\rm e}^\phi\right)
    + \frac{1}{2\pi}\int\limits_{\partial M}\!|dz|\;\kappa_z\phi +\frac{1}{\pi R}\int\limits_{|z| = R}\!|dz|\;\phi
+ 4\log R \ ,
\end{eqnarray}
where $M^R=\{z\in M:|z| \leq R\}$. 
The stationary point of this functional $\varphi(z,\bar z)$ is
given by the formula (\ref{cliouville}). In the case of regular solution it defines on $M$ the
hyperbolic metric ${\rm e}^{\varphi} d^2z$ with geodesic boundaries.

On each hole $H_j$ there exists a unique flat metric with the only singularity at $z_j$
such that the boundary $\partial H_j =
\Gamma_j$ is geodesic and  its length  is  $2\pi \frac{\lambda_j}{\sqrt m}$.
It can be constructed as the pull-back by $\rho_j(z)$ of the metric
${\lambda_j^2\over m\,|\rho|^2}d^2\rho$. 
Its  conformal factor is given by the formula
\begin{equation}
\label{f:definition}
\varphi_j(z,\bar z) =
\log\left[\frac{\lambda_j^2}{m\,\left|\rho_j(z)\right|^2}
\left|\frac{d\rho_j(z)}{dz}\right|^2\right]
\end{equation}
and satisfies the  sewing relations
along the boundary:
\begin{equation}
\label{sewing}
\varphi(z,\bar z) = \varphi_j(z,\bar z),
\hskip 5mm
n^a\partial_a \varphi(z,\bar z) = n^a\partial_a  \varphi_j(z,\bar z)\;\;\;\;{\rm for}\;\; z\in \Gamma_j \ .
\end{equation}
Using the expansion (\ref{rho}) one gets
\begin{equation}
\label{asymptotic}
\varphi_j(z,\bar z) =\log \frac{\lambda_j^2}{m}  -\log|z-z_j|^2
+\frac{c_j}{2\Delta_j}(z-z_j) +
\frac{\bar{c_j}}{2\bar \Delta_j }(\bar z-\bar{z_j})
+ {\cal O}\left(|z-z_j|^2\right).
\end{equation}
We define on $H_j^\epsilon$ the regularized classical action
\begin{equation}
\label{class:reg}
S^\epsilon_{\rm\scriptscriptstyle L}\left[H_j,\varphi_j\right] =
\frac{1}{2\pi} \int\limits_{H_j^\epsilon}\! d^2z\;
  \left(\partial \varphi_j\bar\partial \varphi_j +m\, {\rm e}^{\varphi_j}\right)
    + \frac{1}{2\pi} \int\limits_{\partial H_j^\epsilon}\!|dz|\;\kappa_z \varphi_j
+(\lambda_j^2-1)\log\epsilon \ .
\end{equation}
The  complete classical Liouville action for  hyperbolic singularities  then reads
\begin{eqnarray}
\nonumber
S_{\rm\scriptscriptstyle L}\left[\phi\right]
&=&
4\beta \lim_{\epsilon\to 0}
S^{\epsilon}_{\rm\scriptscriptstyle L}\left[\phi\right] \ , \\
\label{modified2}
S^{\epsilon}_{\rm\scriptscriptstyle L}\left[\phi\right]
 &=&  S^{1/\epsilon}_{\rm\scriptscriptstyle L}\left[M,\phi\right]
+\sum\limits_{k =1}^n 
S^\epsilon_{\rm\scriptscriptstyle L}\left[H_k,\varphi_k\right]\ .
\end{eqnarray}
The action above  differs from the one proposed in
\cite{Hadasz:2003kp} only
by a function of $\lambda_j$. This can be easily verified 
using the indentity
\[
\int\limits_{H^\epsilon_j}\!d^2z\;m\,{\rm e}^{\varphi_j(z,\bar z)}
\; = \; 2\pi\lambda_j^2\left(\log r_j{\rm e}^{-\frac{\vartheta_j}{\lambda_j}} -
\log\epsilon\right) 
\; = \;
-2\pi\lambda_j^2\log\left|r_j^{-1}\frac{d\rho_j}{dz}(z_j)\right|
-2\pi\lambda_j^2\log\epsilon
\]
valid for $\epsilon$ small enough.
The difference
is due to the fact that the line integral in (\ref{class:reg}) extends
over the entire boundary of $H_j^\epsilon$ and not only over 
$\Gamma_j$.

With the help of the sewing relations (\ref{sewing}) one can rewrite  the classical action
$ S^{\epsilon}_{\rm\scriptscriptstyle L}\left[\varphi\right]$
(\ref{modified2})
in the form
\begin{eqnarray}
\label{classical}
 S^{\epsilon}_{\rm\scriptscriptstyle L}\left[\varphi\right]
& = & \;
{1\over 2\pi}\int\limits_M\!\!d^2z\;\left(\partial \varphi \bar\partial \varphi
+m\,{\rm e}^\varphi \right)
+ {1\over 2\pi}\sum\limits_{k =1}^n\int\limits_{H^\epsilon_k}\!\!d^2z\;
\left(\partial\varphi_k \bar\partial\varphi_k +m\, {\rm e}^{\varphi_k}\right)
\\
&&
+{1\over 2\pi}\sum\limits_{k =1}^n\!\!\!\!\!
\int\limits_{\;\;\;\;|z-z_k| = \epsilon}\!\!\!|dz|\;\kappa_z\varphi_k
+\frac{\epsilon}{\pi}\int\limits_{|z| = \epsilon^{-1}}\!|dz|\;\varphi
-\left(4+ \sum\limits_{k =1}^n(1-\lambda_k^2)\right)\log\epsilon   \ .
\nonumber
\end{eqnarray}
It follows from (\ref{cliouville}) and (\ref{f:definition}) that 
\begin{equation}
\label{mdependece}
\varphi = -\log\, m + \ldots\ ,
\hskip 1cm
\varphi_k = -\log\, m + \ldots\ ,
\end{equation}
where the dots denote the $m-$independent terms. This implies that the
integrals in the first line of (\ref{classical}) do not depend on $m.$
The only source of $m$ dependence of the classical action are therefore
the line integrals in (\ref{classical}) and consequently the
$m-$dependent term in $S_{\rm\scriptscriptstyle L}\left[\varphi\right]$
is 
\begin{equation}
\label{mdep:action}
 -{2\beta\over \pi}\log\,m \left(
\sum\limits_{k =1}^n\!\!\!\!\!
\int\limits_{\;\;\;\;|z-z_k| = \epsilon}\!\!\!|dz|\;\kappa_z
+2\epsilon\!\!\!\!\int\limits_{|z| = \epsilon^{-1}}\!|dz|\right)
\; = \;
4\beta(n-2)\log\,m\ .
\end{equation}
As was shown in \cite{Hadasz:2003kp} the classical action
(\ref{classical}) satisfies the Polyakov conjecture
\begin{equation}
\label{Polyakov}
\frac{\partial S_{\rm\scriptscriptstyle L}\left[\varphi\right]}{\partial z_j} 
\; = \; 
-c_j \ .
\end{equation}
Once the accessory parameters $c_j$ are known the relations
(\ref{Polyakov}) allow to compute the classical Liouville action up to
the $z_j-$independent terms. 

We conclude this section calculating the partial derivatives of the classical action (\ref{classical}) with
respect to the parameters $\lambda_j.$
Using the equations of motion
\begin{equation}
\label{eom:f}
\partial\bar\partial \varphi
={\textstyle \frac{m}{2}}{\rm e}^\varphi\ , \hskip 10mm
\partial\bar\partial \varphi_k
\;=\;0\ ,
\end{equation}
the sewing relations (\ref{sewing}), and integrating by parts one  gets
\begin{eqnarray}
\label{computations}
\frac{\partial}{\partial\lambda_j}
 S^{\epsilon}_{\rm\scriptscriptstyle L}\left[\varphi\right]
&=& \frac{1}{2\pi}\sum\limits_{k=1}^n \;
 \int\limits_{H_k^\epsilon}\!d^2z\;
\frac{\partial}{\partial\lambda_j}m\,{\rm e}^{\varphi_k}
+\frac{i}{4\pi}\sum\limits_{k=1}^n\!\!\!\!\!
\int\limits_{\;\;\;\;|z-z_k|=\epsilon}\!\!\!\frac{\partial \varphi_k}{\partial\lambda_j}
\left(\bar\partial \varphi_k\,d\bar z - \partial \varphi_k\,dz\right)
\\
\nonumber
&&
+{1\over 2\pi}\sum\limits_{k =1}^n\!\!\!\!\!
\int\limits_{\;\;\;\;|z-z_k| = \epsilon}\!\!\!\!\!|dz|\;\kappa_z
\frac{\partial\varphi_k}{\partial\lambda_j} + 2\lambda_j\log\epsilon\ .
\end{eqnarray}
It follows from the expansion (\ref{asymptotic}) that
the second and the third term in (\ref{computations})
cancel in the limit. Changing the variables 
$
z = \rho_k^{-1}(\rho)
$ in the first one and using (\ref{f:definition}) one gets 
\begin{eqnarray}
\label{identity}
\frac{1}{2\pi}\int\limits_{H_k^\epsilon}\!d^2z\;
\frac{\partial}{\partial\lambda_j}m\,{\rm e}^{\varphi_k(z,\bar z)}
&  = &
\frac{1}{2\pi}\int\limits_{\rho_k^{-1}(H_k^\epsilon)}
\!d^2\rho \;\frac{\partial}{\partial\lambda_j}
\left(\frac{\lambda_k^2}{|\rho|^2}\right) 
\\
& = &
2\delta_{jk}\Big(\pi(l_j + 1/2) - \vartheta_j-
\lambda_j\log\epsilon\Big)\ ,
\nonumber
\end{eqnarray}
where $\rho_k^{-1}(H_k^\epsilon) = \{ \rho\in \mathbb{C}: \epsilon \, {\rm e}^{\vartheta_k \over \lambda_k} 
\leq |\rho | \leq {\rm e}^{{\pi\over \lambda_j}(l_j + 1/2)}\}$ for $\epsilon$ small enough. 
This finally gives
\begin{equation}
\label{lambdy}
\frac{\partial}{\partial\lambda_j} S_{\rm\scriptscriptstyle L}\left[\varphi\right]
\; = \; 4\beta\lim_{\epsilon\to 0}
\frac{\partial}{\partial\lambda_j} S^{\epsilon}_{\rm\scriptscriptstyle L}\left[\varphi\right]
\; = \; -8\beta\Big(\vartheta_j - \pi(l_j + 1/2)\Big) \ .
\end{equation}
Let us note that the Liouville action
$ S_{\rm\scriptscriptstyle L}\left[\varphi\right]$ does not depend on the signs of $\lambda_j$.
In consequence the l.h.s of (\ref{lambdy}) is an odd function of $\lambda_j$. It follows from (\ref{psis})
that one can always choose $\vartheta_j$ to be an odd function of $\lambda_j$ as well.
Then $l_j$ changes to $-l_j -1$ with the change of sign of $\lambda_j$.
 
\section{Classical solution, $n=3$}

In this section we shall analyze the solution of the $SL(2,\mathbb{R})$ 
monodromy problem for three hyperbolic singularities on the Riemann sphere.
The energy-momentum tensor  with singularities 
located at $z=0,1$ and $\infty$ has the form
\begin{equation}
\label{Tsimple}
T(z) = \frac{\Delta_1}{z^2} + \frac{\Delta_2}{(1-z)^2} +
\frac{\Delta_1 + \Delta_2 - \Delta_3}{z(1-z)} \ .
\end{equation}
Let us consider the systems of solutions of the corresponding Fuchsian equation (\ref{Fuchs})
\begin{equation}
\label{H0}
\Psi^{1}(z) =
\left(\begin{array}{r}
\psi^{1}_+(z) \\
\psi^{1}_-(z)
\end{array}\right)=
\left(
\begin{array}{l}
({i\lambda_1})^{-{1\over 2}}\,{\rm e}^{{{i\over 2}\vartheta_1}}\,
\psi(i\lambda_1,i\lambda_2,i\lambda_3;z)\\
({i\lambda_1})^{-{1\over 2}}\,{\rm e}^{-{{i\over 2}\vartheta_1}}\,
\psi(-i\lambda_1,-i\lambda_2,-i\lambda_3;z)
\end{array}\right)
\end{equation}
where 
\[
\psi(\lambda,\mu,\rho;z) = z^{\frac{1+\lambda}{2}}
(1-z)^{\frac{1-\mu}{2}}\,
{}_2F_1\!\left(\frac{1+\lambda - \mu + \rho}{2}, 
\frac{1+\lambda - \mu - \rho}{2},1+\lambda;z\right)
\]
and ${}_2F_1(a,b,c;z)$ is the hypergeometric function. This system has Wronskian normalized to 1
and its monodromy around $z=0$ is diagonal:
\[
\Psi^1({\rm e}^{2\pi i} z) = \left(
\begin{array}{cc}
-{\rm e}^{-\pi\lambda_1}&0\\
0&-{\rm e}^{\pi\lambda_1}
\end{array} 
\right) \Psi^1(z) \ .
\]
We shall show that for arbitrary hyperbolic weights we can choose a real number
${\vartheta_1}(\lambda_1,\lambda_2,\lambda_3)$ that appears in (\ref{H0})  
in such a way that the monodromy matrix $M^1_2$ at $z=1$,
\[
\Psi^1({\rm e}^{2\pi i} (1-z)) = M^1_2 \, \Psi^1(1- z) \ ,
\]
belongs to $SL(2,\mathbb{R})$. 
To this end let us consider an auxiliary  normalized system
$$
\tilde \Psi^{2}(z) =
\left(\begin{array}{r}
\tilde\psi^{2}_+(z) \\
\tilde\psi^{2}_-(z)
\end{array}\right)=
\left(
\begin{array}{l}
({i\lambda_2})^{-{1\over 2}}\,
\psi(i\lambda_2,i\lambda_1,-i\lambda_3;1-z)\\
({i\lambda_2})^{-{1\over 2}}\,
\psi(-i\lambda_2,-i\lambda_1,i\lambda_3;1-z)
\end{array}\right)
$$
with a diagonal monodromy around $z = 1,$
\[
\label{22monodromy}
\tilde \Psi^2({\rm e}^{2\pi i} (1-z)) = \left(
\begin{array}{cc}
-{\rm e}^{-\pi\lambda_2}&0\\
0&-{\rm e}^{\pi\lambda_2}
\end{array} 
\right) \tilde \Psi^2(1- z) \ .
\]
Using the method of  \cite{Bilal:Gervais} one can calculate the transition matrix 
\[
\Psi^{1}(z) \; = \; S^{\,1\,2}\, \tilde \Psi^{2}(z) \ .
\]
In the case of hyperbolic singularities $\lambda_i \in \mathbb{R}$ it can be written in the form
\begin{equation}
\label{transition}
S^{\,1\,2} = 
i\sqrt{\lambda_1\lambda_2} \left(
\begin{array}{rr}
{\rm e}^{i{\vartheta_1\over 2}} g_-& {\rm e}^{i{\vartheta_1\over 2}} g_+ \\
-{\rm e}^{-i{\vartheta_1\over 2}} \overline{g_+}& -{\rm e}^{-i{\vartheta_1\over 2}} \overline{g_-}
\end{array}
\right)
\end{equation}
where
\[
g_\pm=
\frac{\Gamma(i\lambda_1)\Gamma(\pm i\lambda_2)}
{\Gamma\left(\frac{1+i\lambda_1\pm i\lambda_2+i\lambda_3}{2}\right)
\Gamma\left(\frac{1+i\lambda_1\pm i\lambda_2-i\lambda_3}{2}\right)} \ .
\]
With the formula (\ref{transition}) one can easily calculate
the monodromy $M^1_2$ of $\Psi^{1}(z)$ at $z =1$ :
\begin{eqnarray*}
M^{1}_{2} &=&{S}^{\,1\,2} \left(
\begin{array}{cc}
-{\rm e}^{-\pi\lambda_2}&0\\
0&-{\rm e}^{\pi\lambda_2}
\end{array} 
\right)
\left({S}^{\,1\,2}\right)^{-1} \\
 &=&
\lambda_1\lambda_2\left(\begin{array}{cc}
{\rm e}^{-\pi\lambda_2}|g_{-}|^2 - {\rm e}^{\pi\lambda_2}|g_{+}|^2
& \hspace*{2mm}
-\left({\rm e}^{\pi\lambda_2} - {\rm e}^{-\pi\lambda_2}\right){\rm e}^{i\vartheta_1}
g_{+}g_{-}
\\[10pt]
\left({\rm e}^{\pi\lambda_2} - {\rm e}^{-\pi\lambda_2}\right){\rm e}^{-i\vartheta_1}
\overline{g_{+} g_{-}}
&
{\rm e}^{\pi\lambda_2}|g_{-}|^2 - {\rm e}^{-\pi\lambda_2}|g_{+}|^2
\end{array}\right) \ .
\end{eqnarray*}

Using elementary properties of the gamma function one checks that
$\det S^{\,1\,2} = 1$. Thus  the monodromy matrix $M^1_1$ belongs to $SL(2,\mathbb{R})$
if and only if 
$g_+ g_- {\rm e}^{i\vartheta_1}$ is a real number. This yields the condition :
\begin{equation}
\label{eta}
{\rm e}^{2i\vartheta_1(\lambda_1,\lambda_2,\lambda_3)} = \frac{\overline{ g_+  g_-}}{g_+ g_-} 
\; = \;  {\Gamma^2(-i\lambda_1)\over\Gamma^2(i\lambda_1)}
\frac{
\gamma\left(\frac{1+i\lambda_1 + i\lambda_2 + i\lambda_3}{2}\right)
\gamma\left(\frac{1+i\lambda_1 - i\lambda_2 + i\lambda_3}{2}\right)}
{\gamma\left(\frac{1-i\lambda_1 - i\lambda_2 + i\lambda_3}{2}\right)
\gamma\left(\frac{1-i\lambda_1 + i\lambda_2 + i\lambda_3}{2}\right)}
\nonumber
\end{equation}
where 
\[
\gamma(x) = \frac{\Gamma(x)}{\Gamma(1-x)} \ .
\]
It follows that $\Psi^1(z)$  with 
  $\vartheta_1(\lambda_1,\lambda_2,\lambda_3)$ determined by (\ref{eta}) 
is a normalized solution to the  $SL(2,\mathbb{R})$ monodromy problem 
with singularities at $z=0,1,\infty$
and with a diagonal monodromy at $z=0$. Using the transformation properties of the 
the Fuchsian equation (\ref{Fuchs}) and the conformal map 
\begin{equation}
\label{genproj}
w(z) = \left(\frac{z_2-z_3}{z_2-z_1}\right)\,\frac{z-z_1}{z-z_3}
\end{equation}
one can obtain a normalized solution $\Psi(z)$ for arbitrary locations $z=z_1,z_2,z_3$ of singularities with a  diagonal 
monodromy at $z_1$:
\begin{equation}
\label{transf}
\Psi(z)=\left(\frac{dw(z)}{dz}\right)^{-{1\over 2}}\Psi^1(w(z))   \ . 
\end{equation}
With an appropriate choice of a 
conformal map $w(z)$ and a permutation of the conformal weights one can use this  formula 
to obtain solutions diagonal at other locations.

In particular, for the weights $\Delta_1, \Delta_2, \Delta_3$ at $0,1,\infty$ 
 the solution $\Psi^2(z)$ with a  diagonal monodromy at $z=1$ reads
\begin{equation}
\label{S2}
\psi^2_\pm(z) = (i\lambda_2)^{-{1\over 2}} {\rm e}^{\pm {1\over 2} \vartheta_1(\lambda_2,\lambda_1,\lambda_3)} \ 
(-1)^{1\over 2} \psi(\pm i\lambda_2,\pm i\lambda_1,\pm i\lambda_3,1-z)
\end{equation}
and the solution $\Psi^3(z)$ with a diagonal monodromy at $z=\infty$ is given by
\begin{equation}
\label{S3}
\psi^3_\pm(z) = (i\lambda_3)^{-{1\over 2}} {\rm e}^{\pm {1\over 2} \vartheta_1(\lambda_3,\lambda_2,\lambda_1)}\ 
(-z^2)^{1\over 2} \psi\left(\pm i\lambda_3,\pm i\lambda_2,\pm i\lambda_1,{1\over z}\right)\ .
\end{equation}

\hspace{-10mm}
\includegraphics*[width=.5\textwidth]{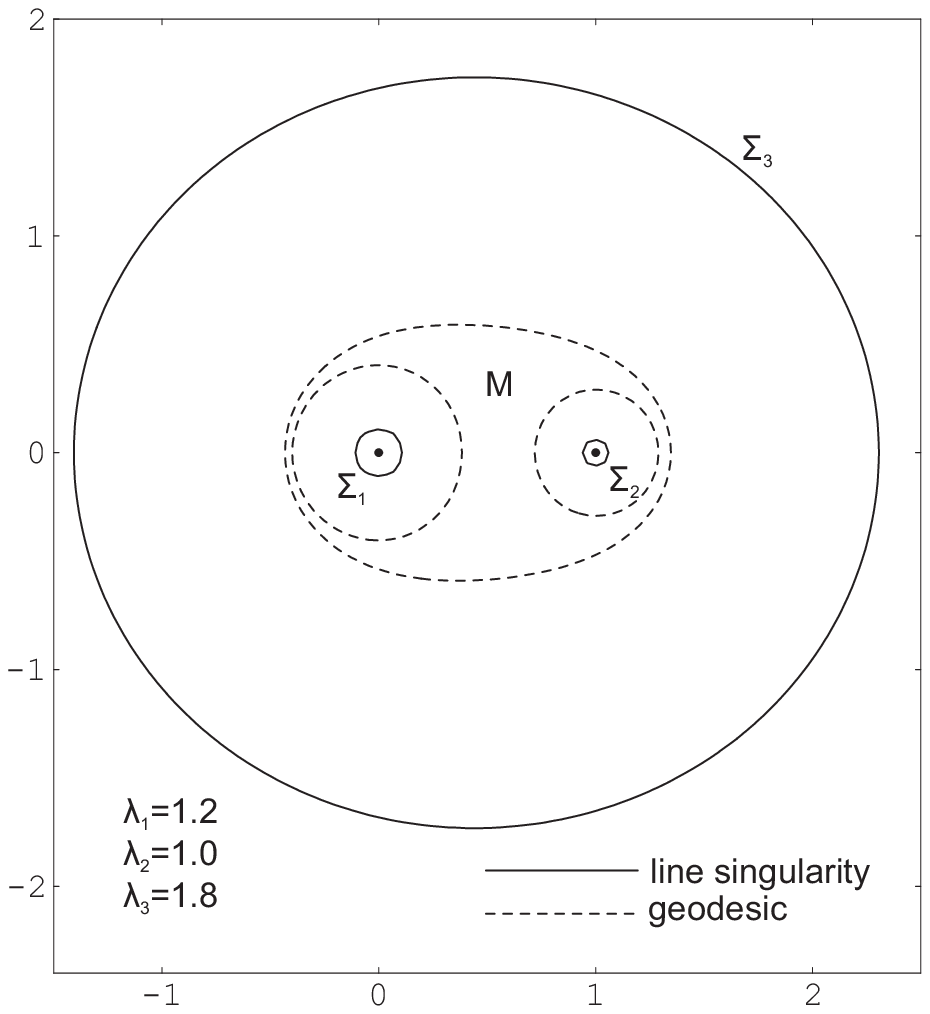}
\includegraphics*[width=.5\textwidth]{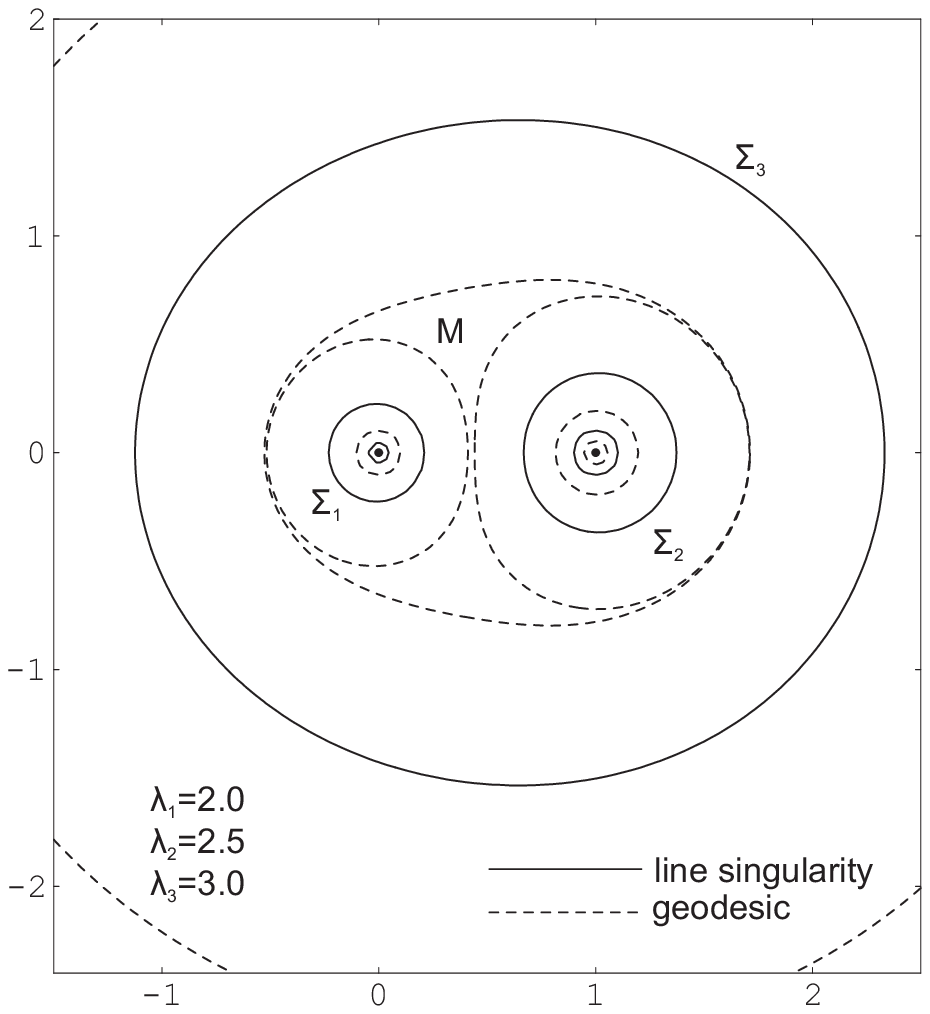}

\vskip 2mm

\centerline{\small{\bf Fig.~1}~ Closed geodesics and line singularities of the metric ${\rm e}^\varphi d^2z$.}

\vskip 3mm

\noindent 

All solutions $\Psi^1(z), \Psi^2(z)$ and $ \Psi^3(z)$ define the same singular solution
to the Liouville equation on the punctured sphere. 
The closed geodesic and the closed line singularities  of the
corresponding hyperbolic geometry ${\rm e}^{\varphi} d^2 z$  
can be found as appropriate levels of the functions $\rho_j$ in regions
where they are invertible. Two examples of this structure 
are presented on Fig.1.

\section{Classical Liouville action, $n=3$}

For arbitrary location of singularities of $T(z)$ the accessory parameters
can be determined from the regularity of the energy--momentum tensor at
infinity or, equivalently, from (\ref{Tsimple})
and the transformation properties of
$T(z)$ applied in the case of the global conformal map (\ref{genproj}).
They have the form
\begin{equation}
\label{acc:3holes}
c_i \; = \; \frac{\Delta_i + \Delta_j-\Delta_k}{z_j-z_i} +
\frac{\Delta_i + \Delta_k-\Delta_j}{z_k-z_i}
\end{equation}
with $i,j,k = 1,2,3,$\ $i\neq j \neq k.$ Integrating the equations (\ref{Polyakov}),
 and their complex conjugate counterparts one then obtains:
\begin{eqnarray}
\label{action:1}
S_{\rm\scriptscriptstyle L}\left[\varphi\right]
& = &
\left(\Delta_1+\Delta_2-\Delta_3\right)\log|z_1-z_2|^2 +
\left(\Delta_2+\Delta_3-\Delta_1\right)\log|z_2-z_3|^2 \\
& + &
\left(\Delta_3+\Delta_1-\Delta_2\right)\log|z_3-z_1|^2
+ R(\lambda_1,\lambda_2,\lambda_3)\ .
\nonumber
\end{eqnarray}
Proceeding to the limit $z_1 \to 0,\, z_2 \to 1, \,z_3 \to \infty$ in (\ref{lambdy})
one gets the equations 
\begin{equation}
\label{R}
{\partial \over \partial \lambda_j}R(\lambda_1,\lambda_2,\lambda_3)=
 -8\beta\Big(\vartheta_j(\lambda_1,\lambda_2,\lambda_3) - \pi(l_j + {\textstyle{1\over 2}})\Big) 
\end{equation}
where $\vartheta_1$ is given by (\ref{eta}), and $\vartheta_2$ and $\vartheta_3$
are determined by (\ref{S2}) and (\ref{S3}), respectively.

In order to calculate the constants $l_j$ one can  consider 
the limit of small positive $\lambda_j$. It follows from (\ref{eta})
that 
$
\lim\limits_{\lambda_j \to 0_+}\vartheta_j(\lambda_1,\lambda_2,\lambda_3) = \pi 
$.
Thus, for $\lambda_j\to 0_+$, 
$$
\rho_j(z) = {\rm e}^{a+{\pi\over \lambda_j} } (z-z_j) + {\cal O}(\lambda_j) (z-z_j) + {\cal O}((z-z_j) ^2) \ ,
$$
where $a$ does not depend on $\lambda_j$. The geodesics around $z_j$ are then described by the equations
$$
|z-z_j| \approx {\rm e}^{-a-\frac{\pi}{\lambda_j}}{\rm e}^{\frac{\pi}{\lambda_j}(l +\frac12)}
\ .
$$
For $\lambda_j$ small enough the ``most distant from $z_j$'' closed geodesic should be still
close to $z_j$. On the other hand the next to the ``most distant'' singular line
should stay away from $z_j$ in this limit. Both conditions are met only for 
$l_j=0$.

Integrating (\ref{R}) and taking into account 
(\ref{mdep:action}) one finally gets
\begin{eqnarray}
R(\lambda_1,\lambda_2,\lambda_3) & = &
 8\beta\!\!\!\!\sum\limits_{\sigma_1,\sigma_2 = \pm}\!\!\!
F\left(\frac{1+i\lambda_1}{2} + \sigma_1\frac{i\lambda_2}{2}
 + \sigma_2\frac{i\lambda_3}{2}\right) 
\nonumber \\
& + &  8\beta\sum\limits_{j=1}^3\Big(H(i\lambda_j) + {\pi\over 2}|\lambda_j|\Big) +4\beta\log m + {\rm const.}
\end{eqnarray}
with 
\begin{equation}
\label{functions:FH}
F(x) \; = \; \int_{\frac12}^x\!dy\; \log\gamma(y),
\hskip 1cm
H(x) \; = \; \int_0^x\!dy\;\log\frac{\Gamma(-y)}{\Gamma(y)}.
\end{equation}

The parameterization of the conformal weights used in
\cite{Zamolodchikov:1995aa} reads
\begin{equation}
\label{Delty1}
\Delta_\alpha \; = \; \alpha(Q-\alpha)
\end{equation}
where $\alpha \in (0,{Q /2}] \subset {\mathbb R}$ or $\alpha \in Q/2 +
 i{\mathbb R}^+.$ The ``background charge'' $Q$ is related to the
 Liouville coupling constant
$b$ as
\[
Q = b + {1\over b}\ .
\]
Comparing the formulae for the central charge obtained in the DOZZ and in
the geometric approach \cite{Takhtajan:yk,Takhtajan:1995fd}
\[
c_{\scriptscriptstyle\rm L} = 1+ 6Q^2 = 1 + 48\beta
\]
as well as the expressions for the conformal weights  (\ref{Delty:geom}) and
(\ref{Delty1}) one obtains the relations
\begin{equation}
\label{relations}
\beta = \frac{Q^2}{8}\ ,
\hskip 1cm
\alpha = \frac{Q}{2}(1+i\lambda)\ ,
\end{equation}
where $\lambda \in {\mathbb R}$ in the case of hyperbolic weights. The
semi-classical regime of the Liouville theory corresponds to   
$c_{\scriptscriptstyle\rm L} \gg 1,$ which can be achieved by taking $b
\to 0.$  Consequently the relations (\ref{relations}) turn into
\begin{equation}
\label{relations2}
\beta = \frac{1}{8b^2}\ ,
\hskip 1cm
\alpha = \frac{1}{2b}(1+i\lambda)\ .
\end{equation}

The DOZZ three-point function has the form \cite{Zamolodchikov:1995aa}
\begin{eqnarray}
\label{DOZZ:3pt}
C(\alpha_1,\alpha_2,\alpha_3) 
& = &
\left[\pi\mu\gamma\left(b^2\right)b^{2-2b^2}\right]^{(Q-\sum\alpha_i)/b}
\times \\
&&
\frac{\Upsilon_0\Upsilon(2\alpha_1)\Upsilon(2\alpha_2)\Upsilon(2\alpha_3)}
{\Upsilon(\sum\alpha_i -Q)\Upsilon(Q+\alpha_1-\alpha_2-\alpha_3)
\Upsilon(\alpha_1+\alpha_2-\alpha_3)\Upsilon(\alpha_1-\alpha_2+\alpha_3)}
\nonumber
\end{eqnarray}
where for $x$ in the strip $0 < \Re(x) < Q$ the function $\Upsilon(x)$
can be defined through the integral representation
\begin{equation}
\label{Upsilon:def}
\log\Upsilon(x) 
\; = \; 
\int\limits_0^\infty\frac{dt}{t}
\left[\left({\textstyle\frac{1}{2}}Q - x\right)^2{\rm e}^{-t}
- \frac{\sinh^2\left({\textstyle\frac{1}{2}}Q  - x\right)\frac{t}{2}}
{\sinh\frac{bt}{2}\sinh\frac{t}{2b}}\right]
\end{equation}
and
\begin{equation}
\label{Upsilon0}
\Upsilon_0 \;=\; \left.\frac{d\Upsilon(x)}{dx}\right|_{x=0}.
\end{equation}
From (\ref{Upsilon:def}) one gets the relations
\begin{eqnarray}
\label{jedynka}
\Upsilon({\textstyle\frac{1}{2}}Q ) &=& 1 \ ,\\
\label{shift}
\Upsilon(x+b) &=& \gamma(bx)b^{1-2bx}\Upsilon(x) \ .
\end{eqnarray}
It follows from (\ref{shift}) that for $b \to 0$
\begin{equation}
\label{diffeq}
\frac{d}{dx}\log\Upsilon(x) \; \approx \; 
\frac{1}{b}\left[\log\gamma(bx) + (1-2bx)\log b\right]
\end{equation}
and 
therefore 
\begin{equation}
\label{Uasymptotic}
\log\Upsilon(x)
\; \approx \;
\frac{1}{b^2}\left[F(bx) +\left(bx(1-bx) -\frac14\right)\log b\right]
\end{equation}
with the integration constant determined from (\ref{jedynka}). One also
checks that for $b\to 0$
\begin{equation}
\label{U0:asympt}
\log\Upsilon_0 \;\approx\; \frac{1}{b^2}\left[F(0) - \frac14\log b\right].
\end{equation}
Using (\ref{Uasymptotic}) and the identity ($\lambda \in \mathbb{R}$) :
$$
F(1+i\lambda) = F(0)- {\textstyle {1\over 2}}\pi  |\lambda| - H(i\lambda)  + i\lambda(\log |\lambda|-1 )  
\ ,
$$
one obtains
\begin{eqnarray}
\label{rel2}
\log\Upsilon\left(\frac{1+i\lambda}{2b}\right)
&\approx&
\frac{1}{b^2}\left[F\left(\frac{1+i\lambda}{2}\right) +
\frac{\lambda^2}{4}\log b\right] \ ,
\\
\label{rel1}
\log\Upsilon\left(\frac{1+i\lambda}{b}\right)
& \approx & 
\frac{1}{b^2}\left[F(0) - H(i\lambda)
- \frac12 \pi|\lambda|
+ \left(\lambda^2 -\frac14\right)\log b\right]
\\
&+&
\frac{i\lambda}{b^2}\left[\log\frac{|\lambda|}{b} -1\right] \ .
\nonumber
\end{eqnarray}
The classical limit of the three-point function \cite{Zamolodchikov:1995aa}
can then be written in the following form
\begin{eqnarray}
\label{3pt:as}
\log C 
& \approx & 
-\frac{1}{b^2}\left[-\;4F(0) + \sum\limits_{\sigma_1,\sigma_2 = \pm}
F\left(\frac{1+i\lambda_1}{2} + \sigma_1\frac{i\lambda_2}{2} +
\sigma_2\frac{i\lambda_3}{2}\right)
\right. 
\nonumber \\
&& 
\left.
\hskip 10mm 
+\;\sum\limits_{j=1}^3\Big(H(i\lambda_j) + {1\over 2} \pi|\lambda_j|\Big) 
+\frac12\log\left(\pi\mu b^2 \right)\right]
\\
&& -\frac{i}{b^2}\sum\limits_{j=1}^3\lambda_j
\left[1-\log |\lambda_j | +
\frac12\log\left(\pi\mu b^2 \right)
\right].
\nonumber
\end{eqnarray}

Under  the reflection  $\alpha \to Q-\alpha$, ($\lambda \to -\lambda$) 
the  DOZZ three point function changes accordingly to 
the formula \cite{Zamolodchikov:1995aa} 
\begin{equation}
\label{reflection}
S(i\alpha_1 - iQ/2) \; = \;
\frac{C(Q-\alpha_1,\alpha_2,\alpha_3)}{C(\alpha_1,\alpha_2,\alpha_3)} \ ,
\end{equation}
where $S$ is the so called reflection amplitude
\begin{equation}
\label{explexpr}
S(P) 
\; = \; 
- \Big(\pi\mu\gamma\left(b^2\right)\Big)^{-2iP/b}
\frac{\Gamma(1+2iP/b)\Gamma(1+2iPb)}{\Gamma(1-2iP/b)\Gamma(1-2iPb)} \ .
\end{equation}
On the other hand the classical Liouville action is by construction
symmetric with respect to this reflection.
This discrepancy can be overcome if one considers instead of $C(\alpha_1,\alpha_2,\alpha_3)$
 the symmetric three-point 
function $\tilde C(\alpha_1,\alpha_2,\alpha_3)$:
\begin{equation}
\label{newC}
\tilde C(\alpha_1,\alpha_2,\alpha_3)=
\left(\prod\limits_{j=1}^3 \sqrt{S\left(i\alpha_j - i\textstyle{{Q\over 2}}\right) }\right)  C(\alpha_1,\alpha_2,\alpha_3)
\ .
\end{equation}
Taking into account the classical limit of the reflection amplitude for real $\lambda$
\begin{equation}
\label{legfactor}
\log {S\left(-\frac{\lambda}{2b}\right)}
\; \approx \;
\frac{2 i}{b^2}\lambda
\left[1-\log|\lambda| + \frac{1}{2} \log\left(\pi\mu b^2\right)
\right].
\end{equation}
and the formulae (\ref{action:1}), (\ref{mdep:action}), and (\ref{3pt:as})
one easily verifies that in the limit $b\to 0$
$$
\log \tilde C \approx -{1\over b^2}  S_{\rm\scriptscriptstyle L}\left[\varphi\right]   + {\rm const.}
$$
where the relation $m=\pi\mu b^2$ is assumed and the constant on the r.h.s. is independent of $z_j$, $\lambda_j$ and $m$.

\section*{Acknowledgements}

L.H. would like to thank L.Takhtajan for enlightening discussions.

\noindent
The work of L.H. was supported by the EC IHP network
HPRN-CT-1999-000161.

\noindent
Laboratoire de Physique Th{\'e}orique is Unit{\'e} Mixte du CNRS
UMR 8627.

\end{document}